\documentclass[final,3p,times,twocolumn]{elsarticle}

\makeatletter
\def\ps@pprintTitle{%
 \let\@oddhead\@empty
 \let\@evenhead\@empty
 \let\@oddfoot\@empty
 \let\@evenfoot\@empty
}

\usepackage{comment}
\usepackage{amssymb}
\usepackage{amsmath}
\usepackage{graphicx}      
\usepackage{subcaption}    
\usepackage{float}         
\usepackage{xcolor}
\usepackage[
    colorlinks=true,       
    linkcolor=magenta,     
    citecolor=magenta,     
    urlcolor=magenta       
]{hyperref}

\usepackage{lineno}
\usepackage{upgreek}
\begin{document}

\begin{frontmatter}



\title{\textbf{Investigating ultra-thin 4H-SiC AC-LGADs for \\ superior radiation-hard timing applications}} 

\author[label1]{Jaideep Kalani} 
\author[label2,label3]{Saptarshi Datta} 
\author[label3,label4]{Ganesh J Tambve} 
\author[label1]{Prabhakar Palni} 

\affiliation[label1]{organization={School of Physical Sciences, Indian Institute of Technology},
            city={Mandi},
            postcode={175005}, 
            state={Himachal Pradesh},
            country={India}}

\affiliation[label2]{organization={School of Physical Sciences, National Institute of Science Education and Research},
            city={Bhubaneswar},
            postcode={752050}, 
            state={Odisha},
            country={India}}

\affiliation[label3]{organization={Homi Bhabha National Institute},
            city={Mumbai},
            postcode={400094}, 
            state={Maharashtra},
            country={India}}

\affiliation[label4]{organization={Centre for Medical and Radiation Physics, National Institute of Science Education and Research},
            city={Bhubaneswar},
            postcode={752050}, 
            state={Odisha},
            country={India}}

\begin{abstract}
The Low Gain Avalanche Diodes (LGADs) are promising particle detectors for timing resolution better than $50$ ps under a high radiation environment. This study investigates n-in-p LGAD architecture, focusing on ultra-thin sensors of thickness less than $50\ \upmu$m using the WeightField2 program. The capabilities of WeightField2 are demonstrated by comparing its  results with irradiation measurements from an FBK LGAD wafer, showing good agreement across unirradiated and neutron-irradiated conditions. This paper presents device simulations in High Luminosity LHC conditions (lifetime integrated fluence $ \mathcal{O} (10^{14})\ \mathrm{n_{eq}~cm^{-2}}$, temperature $ \approx 243\ \mathrm{K} $), and taking into account radiation damage, gain reduction due to fluence, and lattice defects. It is shown that a 20 $\upmu$m thick sensor achieves the best timing performance. Among Silicon (Si), Diamond (C), and 4H-Silicon Carbide (4H-SiC), we found 4H-SiC to be the most promising: it provides the highest gain value for a fixed thickness and gain implant layer configuration, and best retains high charge collection value and timing capability under increasing fluence up to $50\times10^{14}\ \mathrm{n_{eq}~cm^{-2}}$. A time resolution less than 25 ps is reported with different gain implant concentrations for a $20\ \upmu$m 4H-SiC sensor. This work presents the potential of SiC-based LGADs in high-radiation collider environments.
\end{abstract}


\end{frontmatter}



\section{Introduction} \label{Introduction}

The High-Luminosity Large Hadron Collider (HL-LHC) project aims to increase the instantaneous luminosity to 
\(\approx 7.5\times10^{34}\ \text{cm}^{-2}\text{s}^{-1}\)\cite{Schmidt:2016}. These upgrades, including other future high-energy physics experiments (\cite{fcc-ee}, \cite{fcc-hh}, \cite{eic}) seeking larger event statistics, will lead to severe pile-up conditions. The Low Gain Avalanche Diodes (LGADs) offer a timing resolution of approximately \(50\) ps or lower (\cite{ufsd}, \cite{lgad-tr}), along with spatial resolution below \(100\ \upmu\)m (depending on pixel dimensions \cite{DAmen2022}). The LGADs provide an effective solution to mitigate high pile-up scenarios. Therefore, this technology has been qualified for use in the MIP Timing Detector (MTD) detector (\cite{CMS-MIP},\cite{CMS:2019mip}) of CMS and the High Granularity Timing Detector (HGTD) \cite{ATLAS-HGTD} of ATLAS for HL-LHC operations. \par

\begin{figure}[h]
    \centering
    \includegraphics[width=0.48\textwidth]{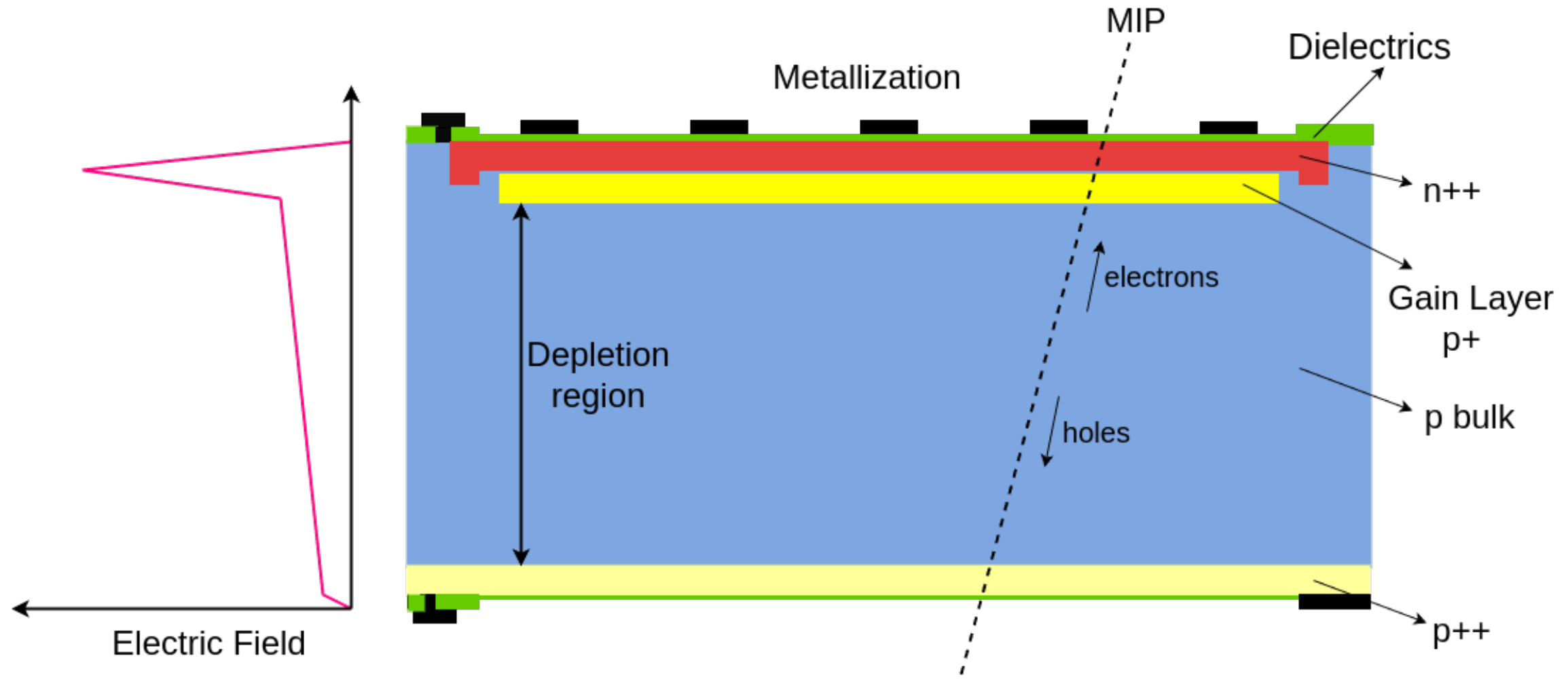} 
    \caption{A typical AC-LGAD architecture}
    \label{fig:1}
\end{figure}

The LGADs outperform standard PIN diodes by introducing an extra gain layer for controlled avalanche multiplication of charge carriers in addition to the drift and diffusion mechanisms. The gain layer is usually a thin, highly doped region with a moderate gain factor (gain \(\sim 10-30\)). This increases the signal amplitude while preserving a high signal-to-noise ratio without the dangers of noise-induced breakdown. A key advancement is the AC-LGAD (AC-coupled) design (Figure \ref{fig:1}), which mitigates DC-LGAD \cite{HotQCD2} (DC-coupled) limitations such as charge-sharing inefficiencies and pixelation constraints due to direct coupling of the readout to the gain layer. In AC-LGADs, the readout is capacitively coupled to the gain layer through an insulating layer; this enables continuous gain layer (increasing fill factor), and improved spatial resolution (much lower than pixel size). These, along with compatibility with modern semiconductor processing techniques, make AC-LGADs suitable for high-granularity, high-resolution timing and tracking (4D) detectors. \cite{Cartiglia:2021b} \par
This study examines the time resolution (\(\sigma_t\)) of an LGAD as a function of operational conditions like bias voltage and radiation damage, including device characteristics like gain implant 
(G.I.) layer concentration, thickness, and bulk material. This work evaluates three different bulk materials, including 4H Silicon Carbide (4H-SiC), where H denotes hexagonal symmetry and a 4-layer per unit cell poly-type of SiC (\cite{DeNapoli:2022} \cite{SiC:2025}). It is well-suited for detector applications due to high saturation drift velocity of carriers, high bandgap (so low intrinsic charge concentration), high breakdown field, and good thermal conductivity. The operational qualities for a 4H-SiC-based LGAD are presented and compared to a Si-based LGAD with similar configurations to demonstrate the former's superiority as a radiation-hard 4D detector. Throughout this study, SiC is used as an alias for the 4H-SiC polytype for brevity.
\par

\subsection{Theoretical framework of the LGAD mechanism}\label{Theory}
The impinging minimum ionising particle (MIP) produces primary $e^-h^+$ pairs while passing through the sensor bulk; these primary carriers further produces cascade of secondary carriers while passing through the gain layer. A calculation using the Shockley-Ramo theorem (\cite{he2001},\cite{sramo},\cite{Sadrozinski2018}) gives the net current (in full signal time) as: $I\propto N_{gen}~q_0~v_{sat}~(G/d)$ \cite{cartiglia2015}. Here $N_{gen}$ is the number of primary charge carriers generated by the MIP, $q$ is the carrier's charge, $v_{sat}$ is the saturation velocity of the charge carrier in the sensor bulk, $G$ is the gain value of the sensor, and $d$ is the thickness of the sensor. This is the key advantage of the LGAD design over a typical semiconductor sensor with a diode-like design. Clearly the signal is amplified from $N_{gen}~q_{0}$ for a typical diode to a larger factor for LGADs. \par
In LGADs, impact ionisation is the crucial mechanism providing the gain factor. The Massey model \cite{massey2006} is an empirical framework used to simulate such a mechanism. The ionisation rate, characterised by the ionisation coefficients \(\alpha_n\) and \(\alpha_p\) for electrons and holes, respectively, follows an empirical relationship:
\[
\alpha_{n,p}(T) = A_{n,p}(T) \exp\left(-\frac{B_{n,p}(T)}{E}\right),
\]
where \(A_{n,p}(T)\) and \(B_{n,p}(T)\) are material-dependent parameters that account for temperature variations, and \(E\) is the local electric field. Furthermore, material-dependent parameter \(B_{n,p}(T)\)  is represented as \(B_{n,p}(T)=C_{n,p}+D_{n,p}\cdot T\). These \(A_{n,p}\), \(C_{n,p}\), and \(D_{n,p}\)\ are six model parameters which are fine-tuned by comparing simulations and empirical data. Corrections dependent on temperature and doping concentration are introduced to make the model more accurate. Comparative TCAD studies (\cite{tcad}, \cite{curras2023}) demonstrate that the Massey model of LGAD mechanism agrees with experimental data across multiple regimes, including gain versus bias voltage, gain versus temperature, and breakdown voltage predictions. We used the Massey model in all simulations and the model parameters are fixed within the WF2 simulation framework as per published dafault values of the Massey model \cite{massey2006}.\par

\subsection{Radiation damage in LGAD}\label{Radiation damage in LGAD}
The radiation damage in an LGAD is broadly of two categories: surface damage and bulk damage \cite{rad-damage}. The former is due to ionisation energy loss; the particle traversing through the sensor ionises the sensor bulk by energy deposition and forms \(\mathrm{e^-/h^+}\)\ pairs. Some of these form at the interface of the sensor bulk and the oxide layer, and cannot recombine properly, leaving trapped charges at the interface.\par

A fraction of the non-ionising energy loss, causes bulk damage to the lattice. The impinging particle, like hadrons, forms Frankel pairs \cite{frenkel}. Some of these pairs get recombined, while others migrate through the lattice and can interact with the other impurities present in the bulk, producing point defects. Furthermore,  these migrating Frankel pairs have enough energy to create further ionisation and atomic displacements. Here the radiation damage is measured in equivalent \( 1 \, \text{MeV} \) neutron fluence \(\left(1 \, \text{MeV}~\mathrm{ n_{\mathrm{eq}}} \, \text{cm}^{-2}\right)\) \cite{ufsd}.\par

Due to radiation exposure, the dopant (acceptors in this case) concentration can be affected by acceptor removal. This process is incorporated into radiation damage simulations. Acceptor removal is a bulk phenomenon caused by radiation-induced defects that deactivate acceptor atoms in the gain layer, leading to a reduction in the effective doping concentration. This effect becomes more pronounced with increasing fluence and results in a substantial decrease in gain, which in turn degrades charge collection and time resolution \cite{Moll:2021}. However, SiC-based LGADs exhibit enhanced radiation tolerance and maintain stable gain performance even at high fluences \cite{Lebedev:2021}.

This study will show how the effects of radiation damage on a sensor can be mitigated by increasing operational bias voltage and maintaining the detector's performance.

\subsection{Time resolution}\label{time_resolution}
The time resolution of an LGAD for a uniform MIP can be given as \cite{cartiglia2015} \cite{siliconSensor2021}:
\[
\sigma_t^2 = \left[\left(\frac{V_{\text{th}}}{S/t_r}\right)_{\text{RMS}}\right]^2 + \sigma_{\text{jitter}}^2 + \sigma_{\text{TDC}}^2
\]
\[
\sigma_{\text{jitter}} = \frac{N}{S/t_r} \quad \text{and} \quad \sigma_{\text{TDC}} = \frac{\text{TDC}_{\text{bin}}}{\sqrt{12}}
\]
where, \( V_{\text{th}} \) is the threshold voltage, \( S \) is the signal amplitude, and \( t_r \) is the signal rise time. These all contribute to the first term in the time resolution formula and are collectively referred to as the time walk term (\( \sigma_{\text{time walk}} \))\cite{cartiglia2014}. The \( \sigma_{\text{jitter}} \) is the jitter or noise term arising in the signal mainly due to electronics, where \(N\) is the rms voltage noise and \(S/{t_r} \) is equivalent to the slew rate at which the detector is performing. \({TDC}_{bin}\) is the least count of the Time-to-Digital-Convertor. We are not using non-uniform charge deposition in this study. Non-uniform charge deposition (Landau fluctuations) provide an intrinsic limit on time resolution. While doing the WF2 simulations, we do not take this into account to deliver conformal statistics without the need to run longer bunch simulations.\par
Moreover, slew rate is  given as the maximum rate of change of a signal \( \left({dV}/{dt}\right) \). Higher slew rates create sharper and more distinct signals. However, there is a trade-off that higher slew rates require a higher amplifier bandwidth, which introduces a higher noise level. The optimal slew rate is finite and electronics dependent \cite{Sadrozinski2018}.

\begin{figure*}[ht]
    \centering
    \begin{subfigure}[t]{0.48\textwidth}  
        \centering
        \includegraphics[width=\textwidth]{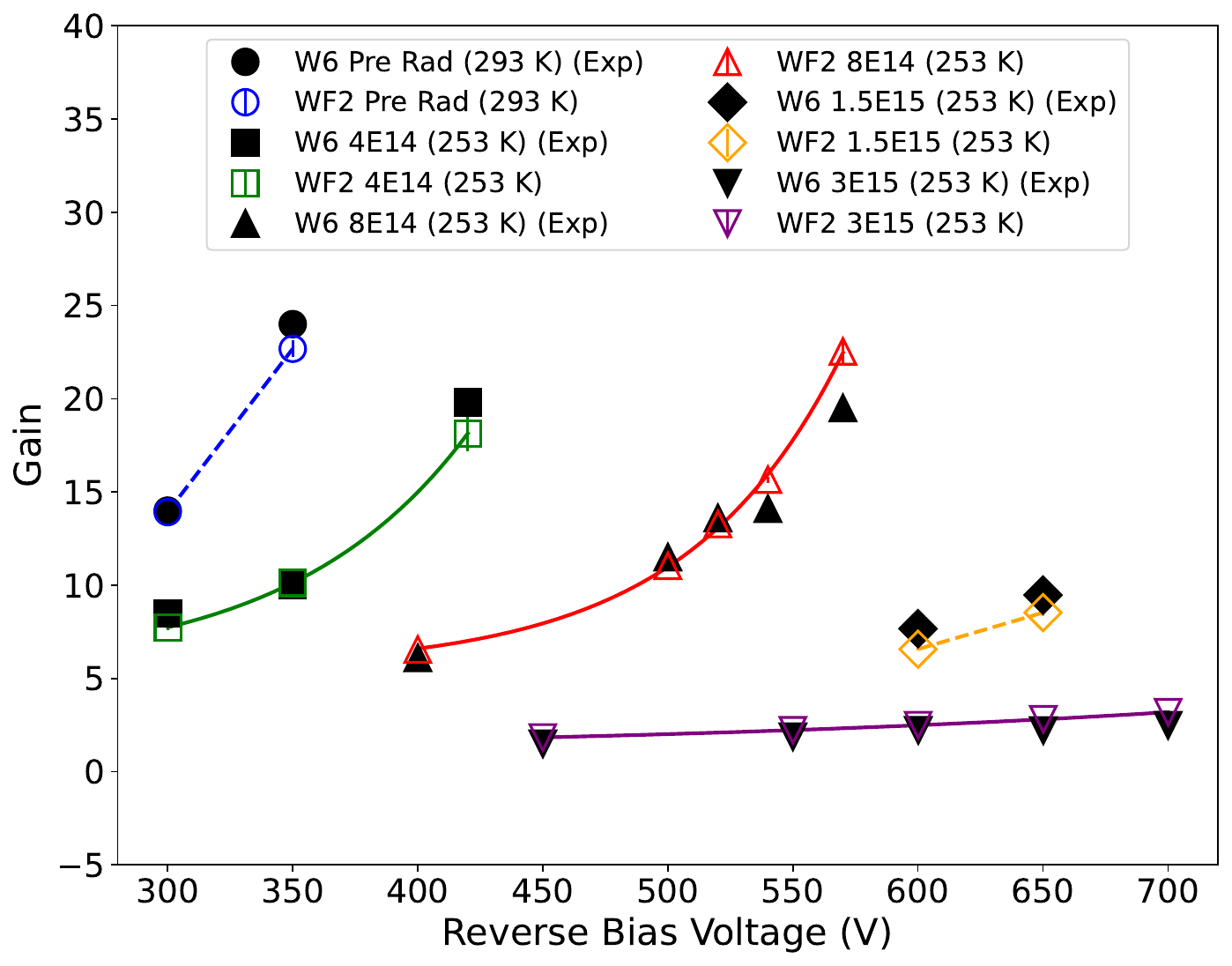}  
        \caption{}
        \label{fig:0a}
    \end{subfigure}
    \hfill
    \begin{subfigure}[t]{0.48\textwidth}  
        \centering
        \includegraphics[width=\textwidth]{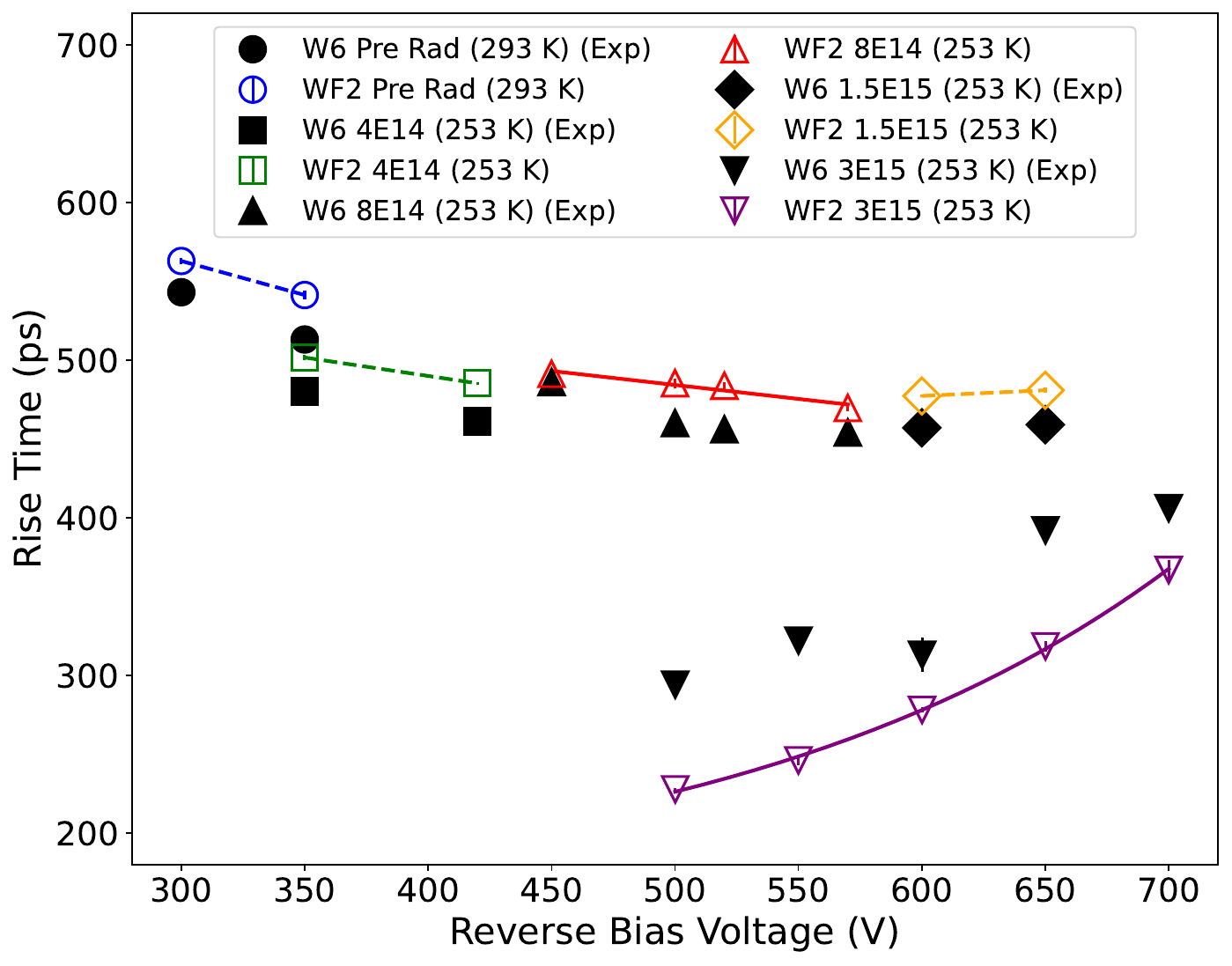}  
        \caption{}
        \label{fig:0b}
    \end{subfigure}
    \caption{Comparison of WF2 simulated (a) gain and (b) rise time  with experimental data from the FBK W6 LGAD sensor \cite{FBK_W6}, for both unirradiated (293~K) and neutron-irradiated (253~K) conditions at various fluence levels. In each plot, the markers labelled as W6 refer to FBK wafer 6 data while markers labelled as WF2 are simulation predictions from WeightField2.}
    \label{fig:0}
\end{figure*}

\section{Simulation methodology}\label{Simulation methodology}
The simulation program \text{WeightField2} (WF2) \cite{cenna2015} build 6.0 is used. It allows us to simulate planar LGADs with possibility of choosing DC or AC functionality. We input the thickness of the device, number of strips, width and pitch of each strip, gain layer doping concentration (in units of $10^{16}$ cm$^{-3}$) and doping profile (depth location of the gain layer doping). For our study we simulated all devices with 
gain implant at $0.5-1.0~\upmu$m from the top. Typically we used 3 or 5 strips with the MIP hitting the middle strip at all cases. The doping concentration of the gain layer is varied from $2.5\times 10^{16}$ cm$^{-3}$ to $5.0\times 10^{16}$ cm$^{-3}$ in some studies, while kept constant in others depending on the study, and is described in the respective sections. The gain mechanism is chosen to be the widely accepted ``Massey LGAD model'' \cite{massey2006}. Typically, the bias voltage is varied with thickness and kept within acceptable operational conditions. It should be noted that for $20~\upmu$m thick silicon devices, the operational bias voltage is limited to 250 V; we have used 300 V (a higher value) in Fig \ref{fig:4} only to qualitatively illustrate the trend \cite{upper-limit}.\par

To run a simulation we choose ''MIP uniform'' impinging (ignoring Landau fluctuations) with 57 / 75 $e^-/h^+$ pairs created per micron, depending on the bulk material \cite{Sciortino2009}. This assumption is standard for the materials studied and allows us to extract the basic features of the signal (analogous to a laser beam study). We always choose neutron irradiation with ''acceptor creation'' turned on, charge capture efficiency (CCE) or trapping coefficients set to \(4.9\times10^{-16}~\mathrm{cm^{2}/ns}\) and \(6.2\times10^{-16}~\mathrm{cm^{2}/ns}\) (standard values) for $e^-$ and $h^+$ respectively. 
We have also studied changes in device properties for change in operational temperature from 243 K (-30$^\circ$C) to 293 K (+20$^\circ$C).  We have provided the simulation conditions for each study in tables provided in the \hyperref[appendix]{Appendix}.\par
Weightfield2 employs a silicon-based gain-layer parametrization, where the gain and bulk regions are treated as parts of the same semiconductor bulk differing only by doping concentration. The same gain-layer parametrization are therefore applied to all bulk materials. Consequently, while the gain layer is modeled identically for silicon, 4H–SiC, and diamond, the 4H-SiC results represent a realistic exploration of bulk transport and signal-shaping characteristics under this imposed gain profile. The diamond results, in contrast, are purely conceptual and serve as a comparison of bulk behavior under an assumed gain mechanism. Similarly, since the charge capture efficiency (CCE) or trapping coefficients values of 4H-SiC are unavailable in literature, we set them same as that of silicon. For comparison between multiple devices we match the gain value of the devices. A plotted line of ``constant gain'' consists of configurations which yield same net charge; this is because we fix the MIP ionisation rate (per micron) in our material and set the gain to be fixed, thus the current is fixed. The acceptor removal mechanism in the gain layer is modelled through the exponential form:
\[
N(\Phi) = N_0~e^{-c\,\Phi}
\]
where \(N(\Phi)\) is the remaining acceptor concentration after irradiation at fluence \(\Phi\), and \(N_0\) is the initial acceptor concentration \cite{Ferrero2019}.

The acceptor removal coefficient \(c\) is not a constant in \textsc{WF2}, rather, it is calculated dynamically in the software and is a function of the initial gain-layer doping type and the irradiation particle type. In the FBK comparison case, in section~\ref{Validation of WF2 predictions with experimental data}, \textsc{WF2} gives \(c = 2.2 \times 10^{-16}~\mathrm{cm}^{2}\), while in the study performed with SiC in Section~\ref{Radiation tolerance of SiC AC-LGADs}, \textsc{WF2} gives \(c = 5.57 \times 10^{-16}~\mathrm{cm}^{2}\). The difference comes from the gain-layer doping type, FBK uses B+C in the gain layer, while in this study B is used as the dopant in the gain layer. More details can be seen in the appendix section for the respective plots.

Moreover, the calculation for time resolution ($\sigma_{t}$) is carried out using a Trans Impedance Amplifier (TI-AMP) for sensor readout at the electronics level, using the standard NA62 model \cite{na62}. The time resolution is calculated according to the formula provided in Section~\ref{time_resolution}. In our implementation, the quantity $V_{\text{th}}$ in the time-walk term is set in the electronics configuration, while the amplitude $S$ and rise time $t_r$ are provided by \textsc{WF2} after each simulation run. The jitter term is also directly reported by \textsc{WF2}, since it depends on the noise and the effective slew rate of the front-end. The TDC bin is set to $20~\mathrm{ps}$ here, following \cite{cartiglia2014}.\par
WF2 gives gain values with two decimal places, time values with 0.01\,ps based on the $Q_{\mathrm{tot}}$ calculation, and $V_{\mathrm{bias}}$ in steps of 1\,V. These should be understood as the granularity of the simulation. The gain implant layer doping concentration is varied up to an order of \(10^{13}\ \mathrm{cm^{-3}}\), which is negligible (0.1\%) compared to the absolute values of the order of \(10^{16}\ \mathrm{cm^{-3}}\). \par

\section{Validation of WF2 predictions with experimental data} \label{Validation of WF2 predictions with experimental data}

To validate the WF2 simulation framework, a comparative study is carried out using results from the FBK W6 wafer sensor \cite{FBK_W6}. This device is a $60~\upmu$m thick silicon DC LGAD sensor from FBK, with dimensions of $1\times1$ mm$^2$. The whole setup was kept at $-27^\circ$C. We simulated a device geometry and operational conditions consistent with the W6 FBK UFSD device ~\cite{FBK_W6} and compared their reported gain and signal rise time values with the WF2 predictions. 

For the FBK W6 device, the exact gain-layer doping concentration is not explicitly reported in the literature. Therefore, it was estimated using the standard (1/$C^2$--$V$) method. In that study, the FWHM of the gain-layer profile is provided, and only the central part of the layer is assumed to be fully depleted. Using this information, a range of possible doping concentrations was calculated from
\[
N_A = \frac{2\,\varepsilon_{\mathrm{Si}}\,V_{\mathrm{GL}}}{q\,t^2}
\]
based on the gain layer thickness~\cite{Ferrero2019}. Here, $t$ is the effective thickness of gain layer, $V_{\mathrm{GL}}$ is the gain-layer depletion voltage, $\varepsilon_{\mathrm{Si}}$ is the permittivity of silicon, and $q$ is the elementary charge. Considering also that \textsc{WF2} can simulate gain-layer thicknesses up to 0.5~\textmu m, a peak initial acceptor concentration of $4.71\times10^{16}$~cm$^{-3}$ was chosen at zero fluence as a representative value for the simulations, as it provides the best agreement with the FBK reported gain and signal rise time values. The active layer thickness was set to 55~$\upmu$m instead of 60~$\upmu$m, as several studies on FBK W6 wafers indicate that the effective active thickness is about 55~$\upmu$m after thermal bonding to the substrate~\cite{FerreroThesis}.

Other structural and operational parameters (e.g.\ number of strips and temperature) were taken from the reference study and kept fixed for consistency. The  configuration for this simulation is provided in appendix section table~\ref{fbk_table}. \par

As shown in Figure~\ref{fig:0a}, the gain (coloured markers) closely aligns with the reported experimental measurements (black markers), both demonstrating a comparable exponential increase with bias voltage. Furthermore, the expected reduction in gain with increasing fluence, along with gain recovery at higher bias voltages, is well reproduced by the simulation. A similar comparison is carried out for the signal rise time as a function of bias voltage. The rise time is defined as the time between 10$\%$ and 90$\%$ of the peak signal amplitude (same as CFD technique used in published data). As can be seen in the Figure~\ref{fig:0b}, trend of decreasing rise time with increasing bias is predicted correctly by WF2.\par

The WF2 measurements shown in Figure~\ref{fig:0a} and Figure~\ref{fig:0b} are fitted, except for the data with only two points, which are connected by a dashed line. The overall agreement of WF2 predictions with published sensor data confirms the capability of WF2 to reliably capture key features of the internal charge multiplication mechanism in LGADs.

\section{Results and discussion}
 We investigated different material choices for the sensor bulk followed by a study on preferred thickness. Further, we present detailed characterisation of the detector's electrical behaviour, signal response, performance metrics under various operating conditions, including high radiation dosage.

\subsection{Choice of sensor bulk material}\label{mat}
\begin{figure*}[ht]
    \centering
    \begin{subfigure}{0.48\textwidth}  
        \centering
        \includegraphics[width=\textwidth]{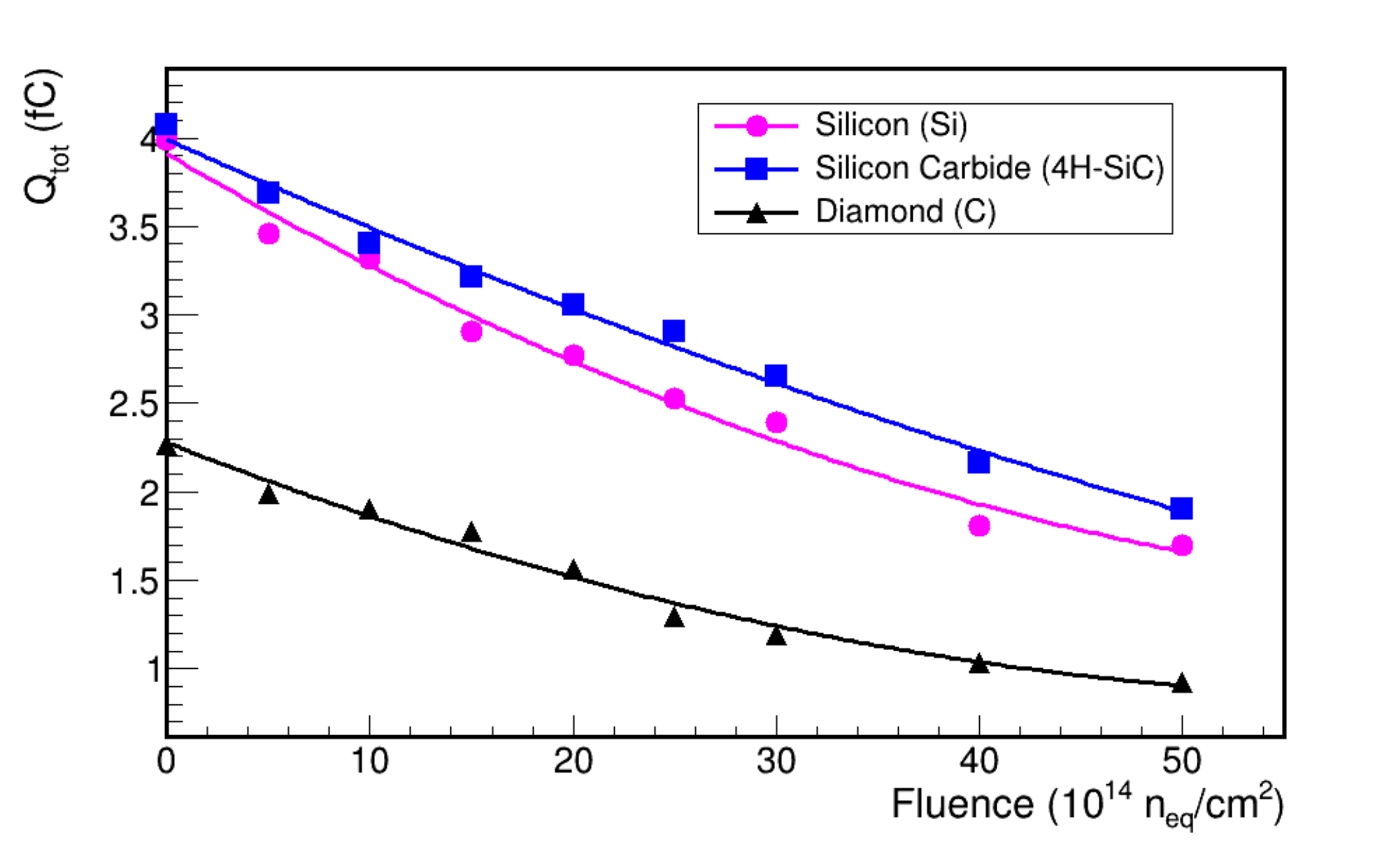}  
        \caption{Charge collected from the LGAD with the TIA decreases with increasing radiation dosage.}
        \label{fig:2a}
    \end{subfigure}
    \hfill
    \begin{subfigure}{0.48\textwidth}  
        \centering
        \includegraphics[width=\textwidth]{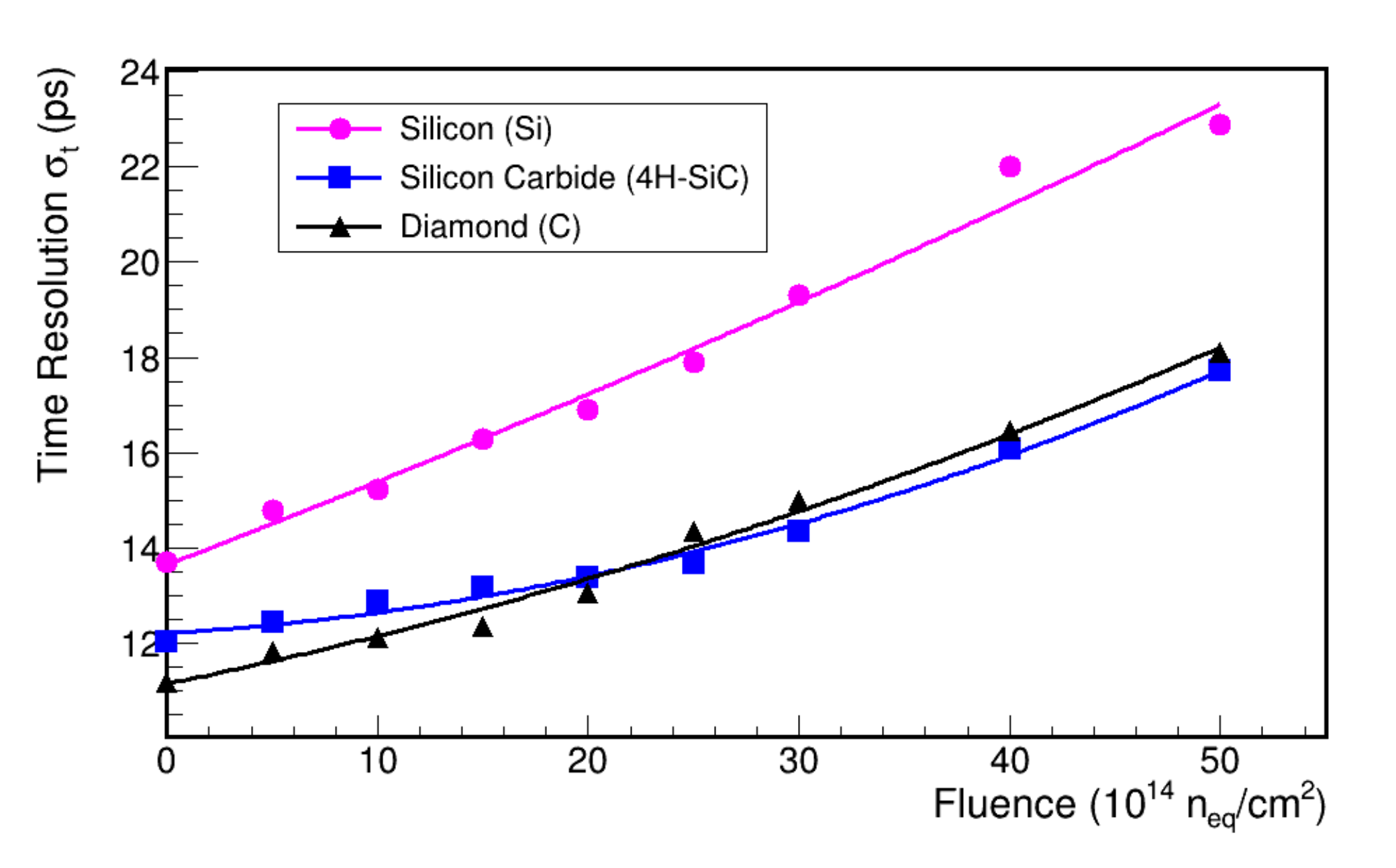}  
        \caption{The timing capability of detectors worsens with increasing radiation dosage.}
        \label{fig:2b}
    \end{subfigure}
    \caption{Simulation results for detectors of \(20~\upmu \)m\ thickness and operated at \(V_{bias}\ =\ 150\ V\)\ and temperature of 243 K.}
    \label{fig:2}
\end{figure*}

We investigated devices with bulk materials Silicon (Si), 4H-SiC, and Diamond. We noticed a significant performance difference in the ultra-thin range (thickness \(20~\upmu\text{m}\)). The bias voltage is kept constant, i.e., 150 V for all three devices while the G.I. concentration is varied to match the gain value of unirradiated devices. Then, for increasing fluence, the variations in both the total charge collection and the time resolution of the devices are studied which is shown in Figure \ref{fig:2}. \par

The choice of material is based on two criteria: maximal total charge collection ($Q_{tot}$) and retaining good time resolution ($\sigma_t$) over an extensive range of fluence, as expected for a tracking detector placed close to the interaction vertex. In Fig~\ref{fig:2a}, the silicon device and the SiC device have similar ${Q_{{tot}}}$ values while the Diamond device performs poorly: at the fluence of $50~\times10^{14}~\mathrm{n_{eq}~cm^{-2}}$ , Si and SiC devices have $\sim2\ \mathrm{fC}$ charge collection while Diamond provides less than $1\ \mathrm{fC}$. In Fig~\ref{fig:2b}, the Diamond device and SiC device have similar $\mathrm{\sigma_t}$ values: they hold a time resolution less than $18\ \mathrm{ps}$, at fluence level of $50\times10^{14} \mathrm{n_{eq}~cm^{-2}}$, whereas the silicon device's time resolution steeply rises to $\sim23~\mathrm{ps}$ at the same fluence level. Therefore, the results clearly show that only the 4H-SiC material performs well, satisfying both the criteria, making it the best choice for sensor bulk material (\cite{Zhao:2024}, \cite{Rafi:2020}).\par
Diamond generates far fewer electron-hole pairs than silicon, resulting in lower total charge (${Q_{tot}}$), but its exceptional radiation hardness, negligible leakage current, and fast carrier transport yield very short rise times and low noise. Because the jitter term scales as $N/(dV/dt)$, the combination of small $N$ and large $dV/dt$ compensates for the reduced signal amplitude, maintaining timing performance comparable to silicon. The superior timing of 4H-SiC relative to silicon arises from its higher carrier drift velocity at operating fields, larger breakdown field enabling stronger bias.

\begin{figure}[h]
    \centering
    \includegraphics[width=\linewidth]{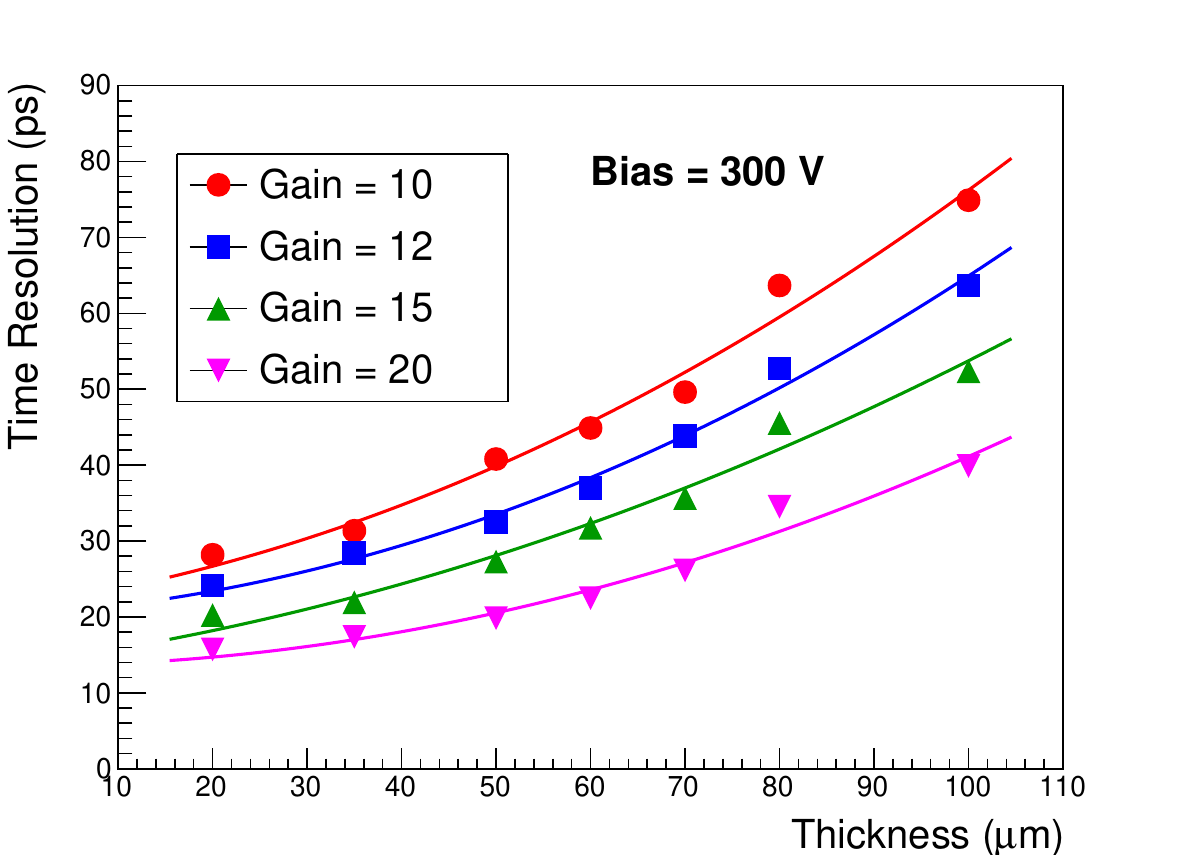}
    \caption{Simulation predictions of time resolution variation as a function of thickness for different fixed-gain SiC bulk AC-LGAD at the constant $V_{bias}$ of $300\ V$}
    \label{fig:3}
\end{figure}
\begin{figure*}[ht]
    \centering
    \includegraphics[width=1.0\textwidth]{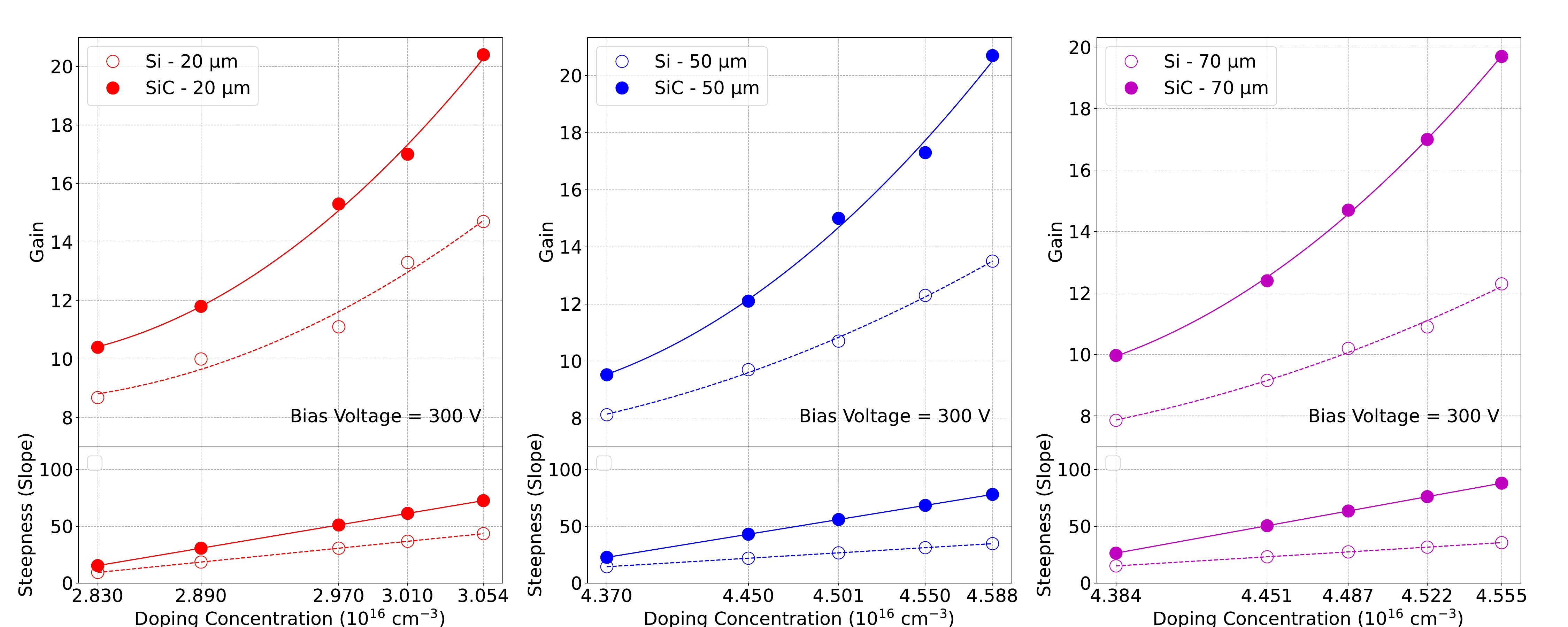} 
    \caption{Figures show WF2 predictions of change in gain value of Si and SiC devices with increasing G.I. dopant concentration. \textit{Top:} The variation of gain with increasing doping concentration of the gain layer in 20 $\upmu$m, 50 $\upmu$m and 70 $\upmu$m SiC and Si bulk AC-LGAD with quadratic fits. \textit{Bottom:} First derivative of the quadratic fits of the corresponding curves.}
    \label{fig:4}
\end{figure*}

\subsection{Thickness dependence of AC-LGAD performance}

The simulations are performed for SiC bulk unirradiated AC-LGAD at a constant temperature ($243\ K$) and bias voltage (300 V), and varying thickness. We increased the G.I. doping concentration with increasing thickness to ensure that in unirradiated condition devices of all thickness provide the same gain value. This allowed us to study the dependence of time resolution on sensor thickness as shown in Figure~\ref{fig:3}.\par

The reduction in the detector thickness from 100 $\upmu$m to 20 $\upmu$m  results in a relative improvement in time resolution by $\sim$ 60\%,  for various gain values. This indicates that using thinner sensor, within technological limits, can significantly benefit the timing performance of LGADs.\par

The timing resolution in LGADs is largely determined by the slew rate, which reflects how fast the signal rises. Even though the charge carriers drift at saturated velocity in all sensor thicknesses, the induced current scales as $G/d$, as discussed in section~\ref{Theory} and section~\ref{time_resolution}.  Hence, for a fixed gain, higher thicknesses results in a reduced $G/d$, leading to a lower slew rate and inferior timing resolution. \par

Figure~\ref{fig:4} (top panel) shows the gain as a function of gain layer doping concentration for silicon and silicon carbide bulks, fitted with quadratic curves. The Figure~\ref{fig:4} (bottom panel) shows the first derivative of the fits, highlighting the gain sensitivity with respect to doping concentration. Notably, the slope for the SiC bulk is steeper and increases more rapidly with doping concentration as compared to Si. This indicates a higher gain sensitivity in SiC under fixed thickness and bias conditions.
An increase in the gain layer doping concentration enhances the probability of impact ionisation, thereby leading to a higher charge multiplication, i.e. gain. This helps in improving the timing capabilities of the detector. Across all thicknesses, SiC consistently exhibits higher gain than Si, indicating superior charge multiplication efficiency.\par
In addition, in fig~\ref{fig:4}, within the considered operating conditions the thickness dependence largely saturates as we go to higher thicknesses, resulting in only minor differences in the G.I.\ layer doping concentration between the 50 and 70~$\upmu$m sensors for achieving the same gain, while the 20~$\upmu$m device shows a lower effective G.I.\ doping concentration for achieving the same gain at the same bias.

 \begin{figure}[h]
    \centering
    \includegraphics[width=\linewidth]{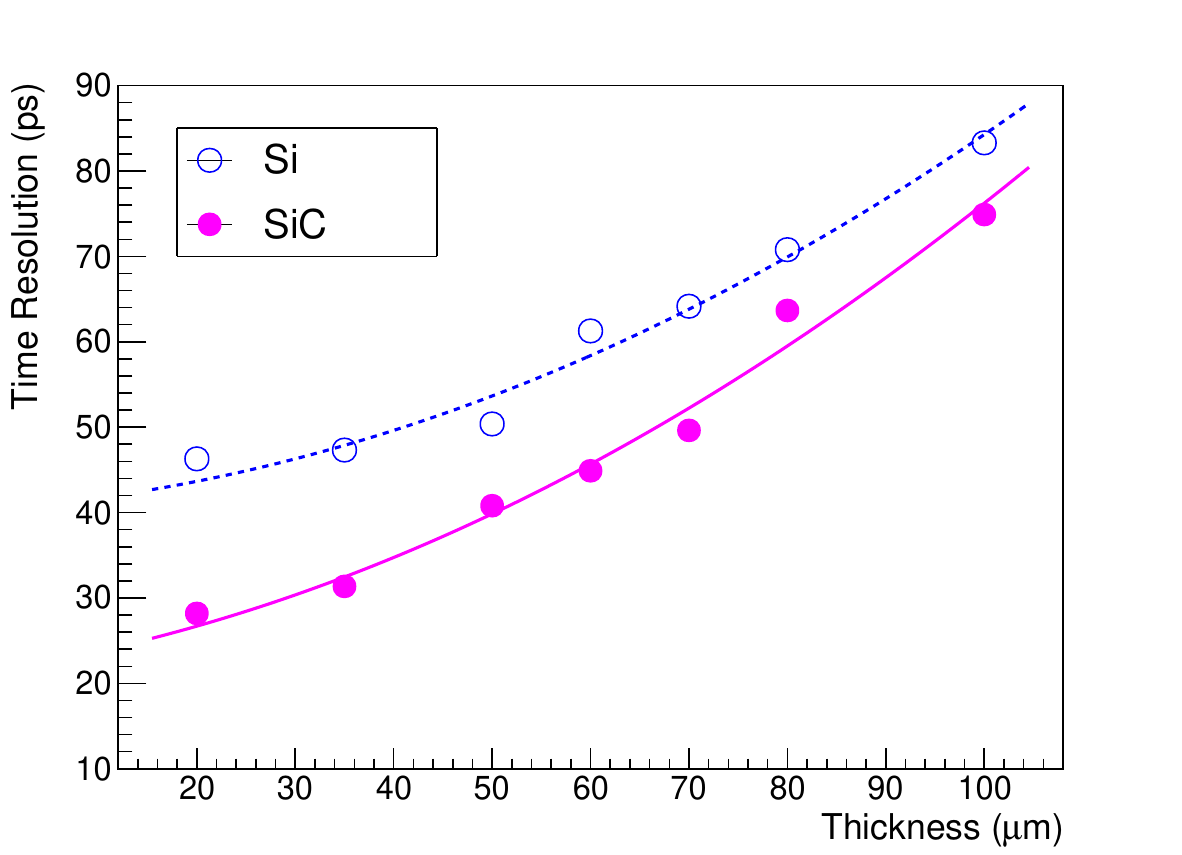}
    \caption{The comparison of simulated time resolution of SiC and Si-bulk AC-LGAD over a thickness range at a fixed gain value of 10. The detectors are operated at fixed bias voltage of 300 V at a temperature of 243 K.}
    \label{fig:5}
\end{figure}

A comparative study of the performance of the SiC bulk detector relative to the Si bulk is shown in Figure~\ref {fig:5}, at gain = 10, and under identical physical conditions. The Si bulk AC-LGAD exhibits an improvement of 44\% in time resolution, while that of the SiC bulk demonstrates an improvement of 60\% while decreasing the thickness from 100\,$\upmu$m to 20\,$\upmu$m.

\subsection{Temperature-dependent measurements}

The AC-LGADs are highly sensitive to variations in temperature, which can significantly impact their time resolution, charge collection efficiency, and gain characteristics \cite{Crowell:1966}. 
The readout electronics generates significant heating (\cite{Whitmore:2010}, \cite{Boetti:2003}) of the sensor. As a result, AC-LGADs experience an increased leakage current due to the localised increase in temperature. The increased leakage current in turn leads to temperature increase. Besides this, there are contributions from shot noise and the Johnson noise; the latter varies as $V_{\text{rms}} \propto \sqrt{T}$, where $V_{\text{rms}}$ is the r.m.s. voltage due to the noise in the detector \cite{johnson1928}.\par

The following measurements present detailed results for the temperature's impact on crucial sensor properties like gain and time resolution. These insights will help to assess the thermal stability of AC-LGADs and thus guide the design of future detectors, optimised for specific operating conditions.

\subsubsection{Effect of  temperature on  gain}
\begin{figure}[h]
    \centering
    \includegraphics[width=\linewidth]{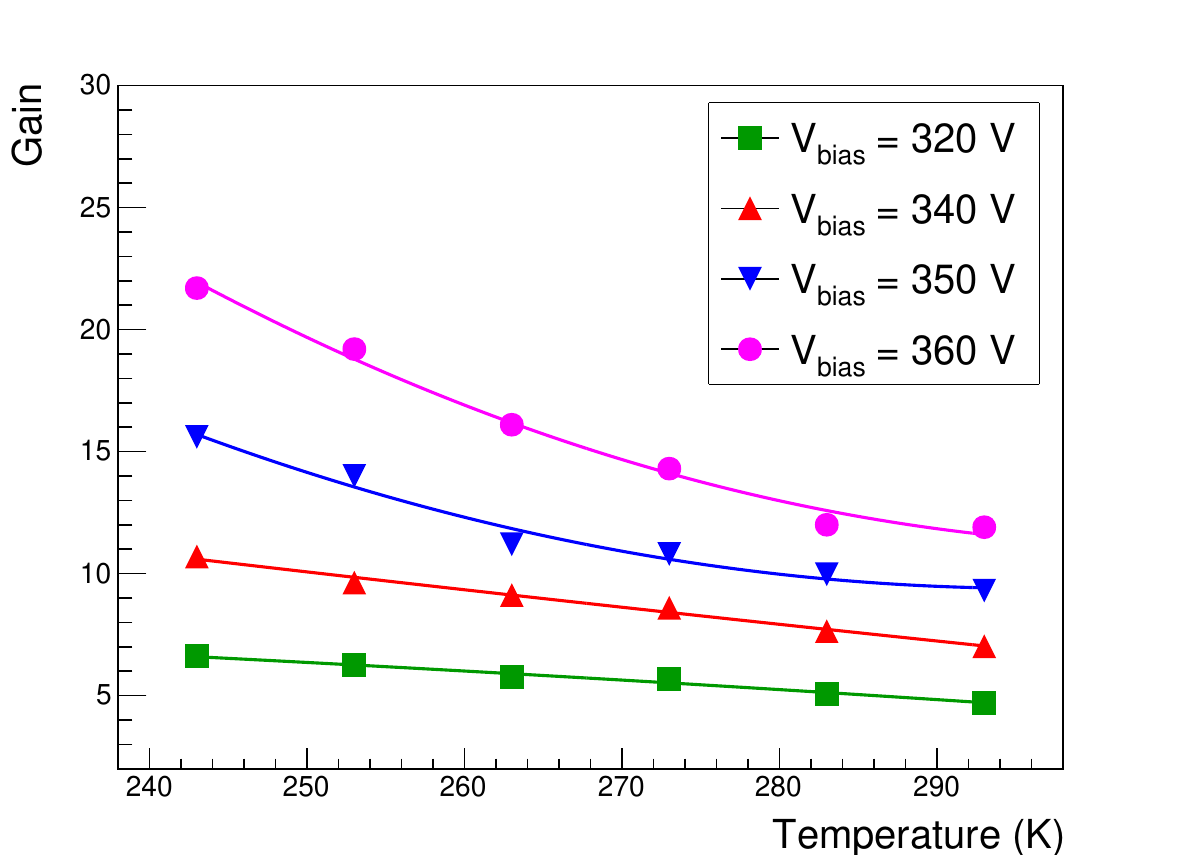}
    \caption{Simulated gain of 20 $\upmu$m SiC bulk AC-LGAD as a function of temperature for different bias voltages.}
    \label{fig:6}
\end{figure}
Figure \ref{fig:6} illustrates the gain in a SiC bulk (20 $\upmu$m)  as a function of temperature. The gain implant doping concentration is set  to  2.301 $\times$ 10$^{16}$ cm$^{-3}$. The data shows gain profiles at four different bias voltages: 320 V, 340 V, 350 V and 360 V. The result indicates a consistent  decrease of gain with the increase in temperature, which agrees with the predictions of the Massey model \cite{massey2006}.
Therefore, AC-LGADs need to operate at higher \(V_{bias}\) with increasing temperature to counteract the decrease in gain value.

\begin{figure*}[ht]
    \centering
    \begin{subfigure}{0.48\textwidth}  
        \centering
        \includegraphics[width=\textwidth]{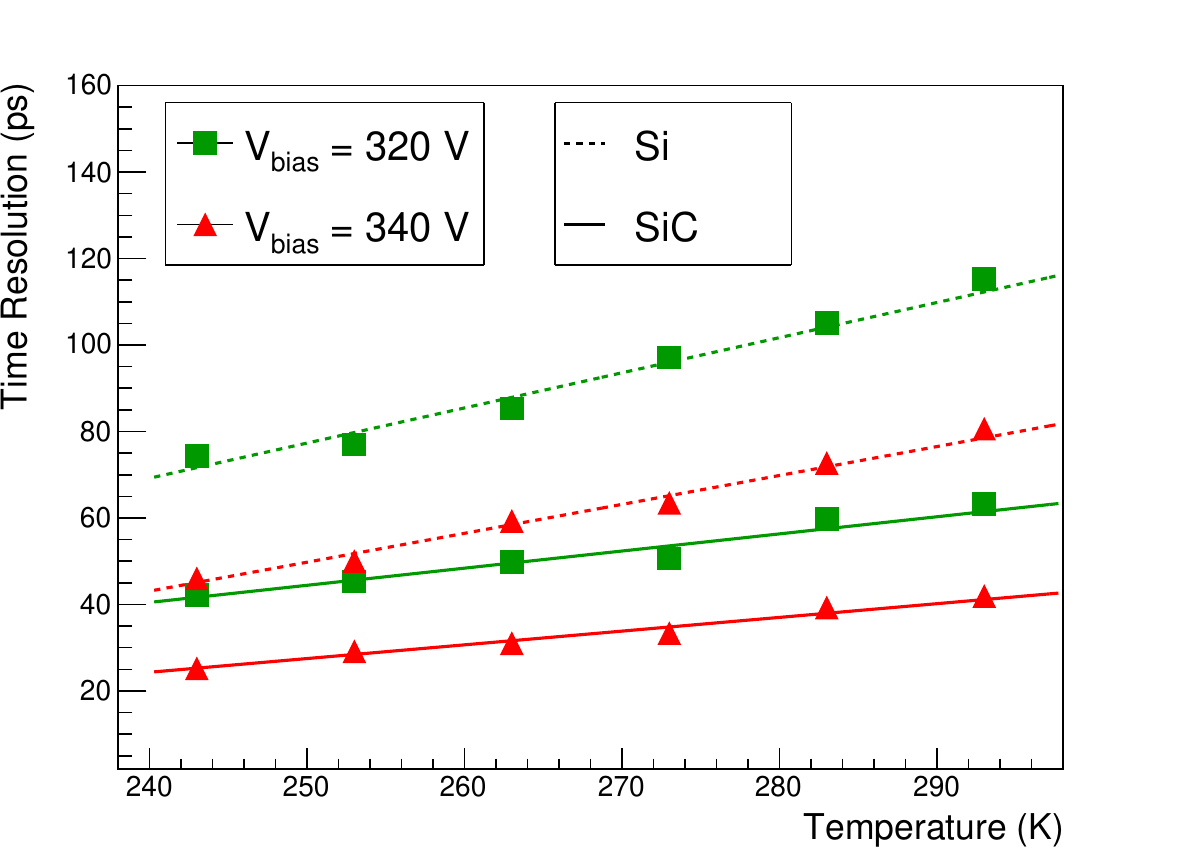}  
        \caption{}
        \label{fig:7a}
    \end{subfigure}
    \hfill
    \begin{subfigure}{0.48\textwidth}  
        \centering
        \includegraphics[width=\textwidth]{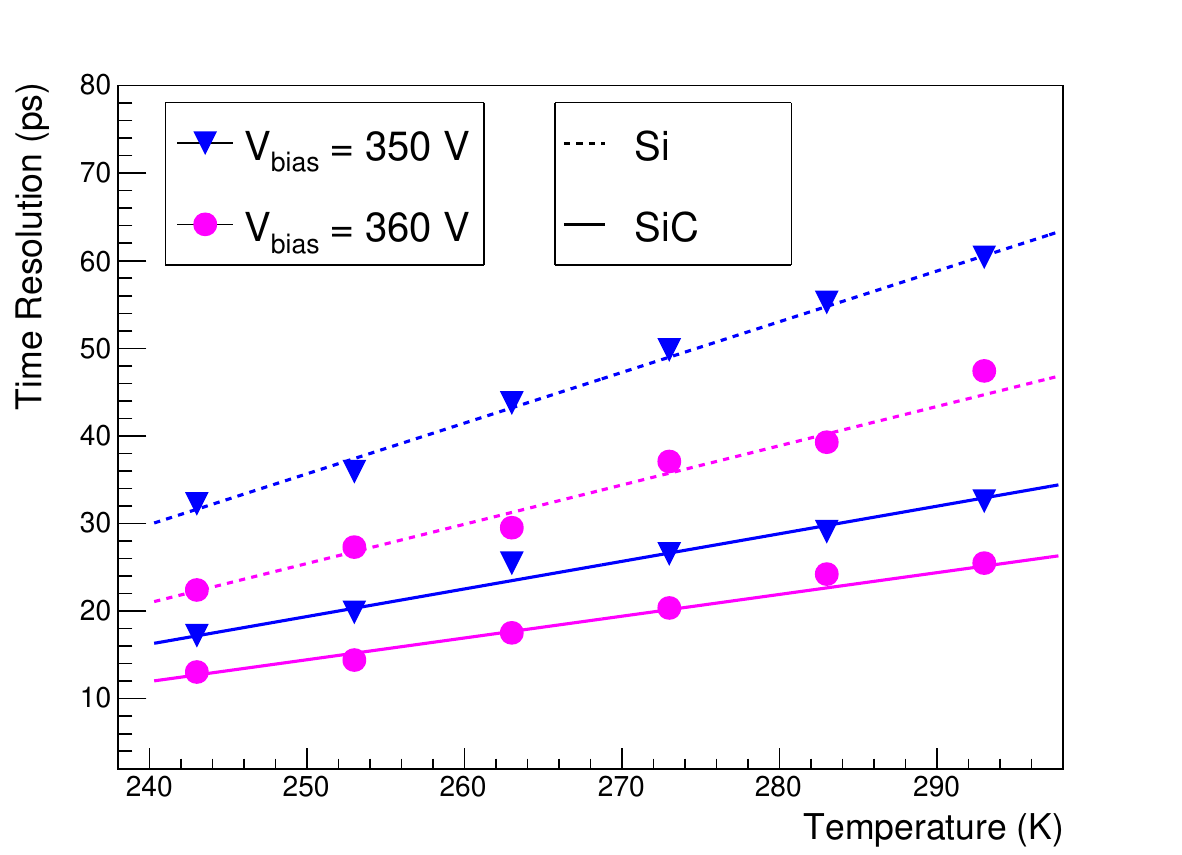}  
        \caption{}
        \label{fig:7b}
    \end{subfigure}
    \caption{WF2 prediction of timing performance comparison between 20 $\upmu$m SiC and Si bulk AC-LGAD with increasing temperature at (a) 320 and 340 V and (b) 350 and 360 V. }
    \label{fig:7}
\end{figure*}

\subsubsection{Performance of time resolution with temperature}

Figure \ref{fig:7} shows the time resolution for a SiC bulk AC-LGAD (G.I. layer doping concentration is 2.301 $\times$ 10$^{16}$ cm$^{-3}$), in solid lines, against the temperature. The data indicates a linearly increasing trend in the time resolution with temperature, as temperature increased from  243 K to 293 K. For comparison, the study is also conducted for silicon bulk AC-LGAD with the same physical conditions. For every operating bias voltage, the SiC device outperforms the Si device over the entire temperature range.
The SiC AC-LGAD shows a  decrease in the time resolution by $\sim$  33\% and $\sim$ 48\% at bias voltages 320 V and 360 V, respectively, when the temperature is decreased from 293 K to 243 K.

In contrast, the Si bulk AC-LGAD exhibits a  decrease in time resolution by $\sim$ 35\% and $\sim$ 52\% at bias 320 V and 360 V, respectively. A marginal reduction in the percentage drop of time resolution in SiC, directly indicates its stability towards the extreme temperature changes. 
Furthermore, the best time resolution achieved is $\sim$ 13 ps at 243 K for SiC and $\sim$ 22 ps for Si, both at a bias voltage of 360 V. This clearly indicates better temperature stability of SiC, as compared to Si. The higher thermal conductivity of SiC plays a crucial role here; it allows the dissipation of heat evenly, preventing any localised heating, which makes it less sensitive towards temperature changes as shown in figure \ref{fig:7}. Moreover, a higher bandgap of  SiC also supports  in reducing leakage current at elevated temperatures.\cite{Rafi:2020}

\begin{figure} 
    \centering
    \includegraphics[width=0.98\linewidth]{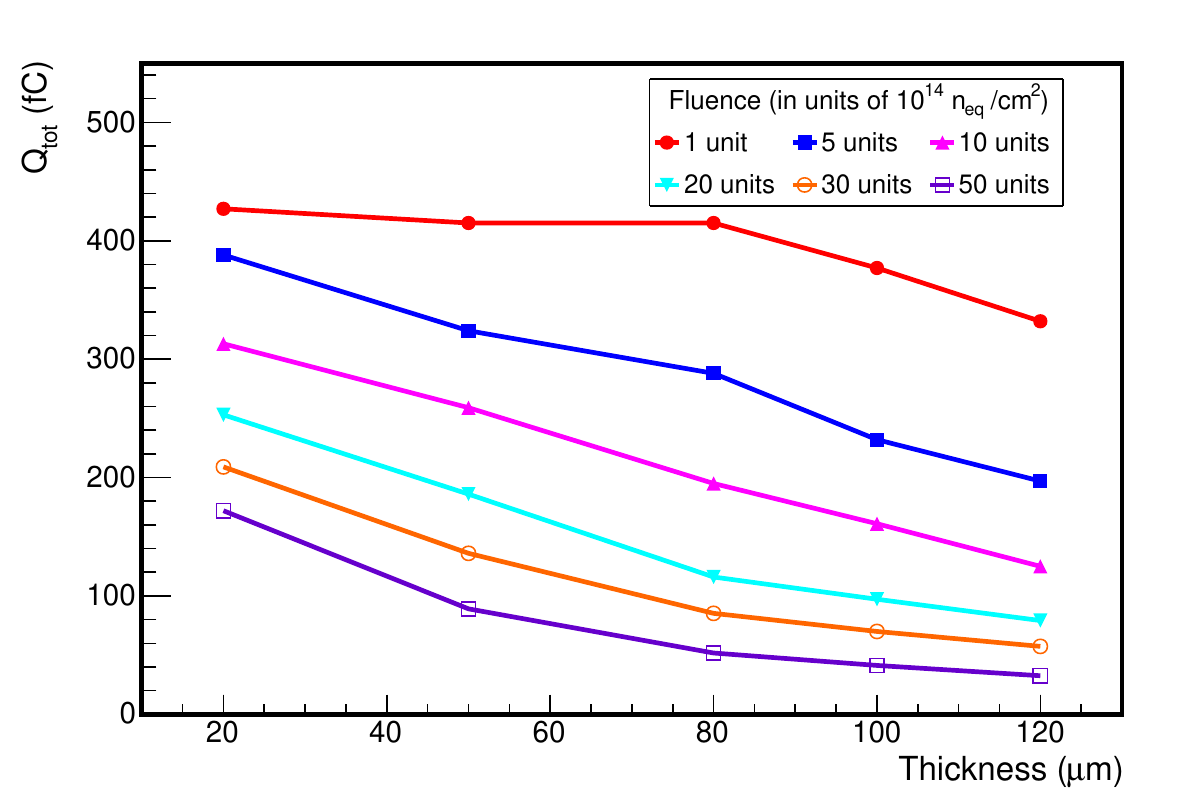}
    \caption{Simulation predictions of variation of \(Q_{tot}\)\ for different sensor thickness with increasing radiation damage. Temperature is set at 253 K and fluence units are in \(10^{14}\ \mathrm{n_{eq}~cm^{-2}}\).}
    \label{fig:8}
\end{figure}

\subsection{Radiation tolerance of SiC AC-LGADs} \label{Radiation tolerance of SiC AC-LGADs}

Radiation dosage received by the detector is expressed in terms of fluence, in the unit of 1 MeV $\mathrm{n_{eq}~cm^{-2}}$. To investigate radiation hardness, SiC AC-LGAD sensor is subjected to various fluence levels, ranging from unirradiated (zero fluence) to as high as \(50 \times 10^{14}\ \mathrm{n_{eq}~cm^{-2}}\). The study is performed for $20~\upmu\text{m}$ sensor at a constant temperature of 253 K (-20$^\circ$C) to mimic the beam-pipe temperature \cite{CMS:2019mip}. For the thicknesses 20 $\upmu$m, 50 $\upmu$m, 80 $\upmu$m, 100 $\upmu$m and 120 $\upmu$m we had the bias voltages as 115 V, 140 V, 142 V, 190 V and 202 V, and the G.I. concentration as (in units of $10^{16}$ cm$^{-3}$) as 4.40, 4.82, 4.81, 4.92 and 4.94 respectively. We chose such values to make sure that at Fluece = 0, detectors of all thickness have gain value of 15. Figure \ref{fig:8} shows that the \(Q_{tot}\)\ drops with an increase in fluence levels, for all the sensor thicknesses. This highlights the adverse effects of radiation damage. Radiation damage reduces carrier transport by introducing lattice defects in the bulk, and, reduces gain by neutralising dopants in the G.I. layer \cite{Raskina:2022}.\par
For a particular thickness, with increasing fluence the main reason for reduced charge is the loss of gain due to acceptor removal in the gain layer. Furthermore, the decrease in total collected charge $Q_{tot}$ with increasing thickness (but fixed fluence) is explained by taking into account bulk defects which cause trapping, recombination, finite lifetime, and incomplete depletion. In ideal, defect-free detectors, $Q_{tot}$ always increases with thickness. However with irradiation, bulk defects increase with thickness and thus the $Q_{tot}$ decreases.

\subsubsection{Fluence-dependent gain in AC-LGADs}
\begin{figure}
    \centering
    \includegraphics[width=\linewidth]{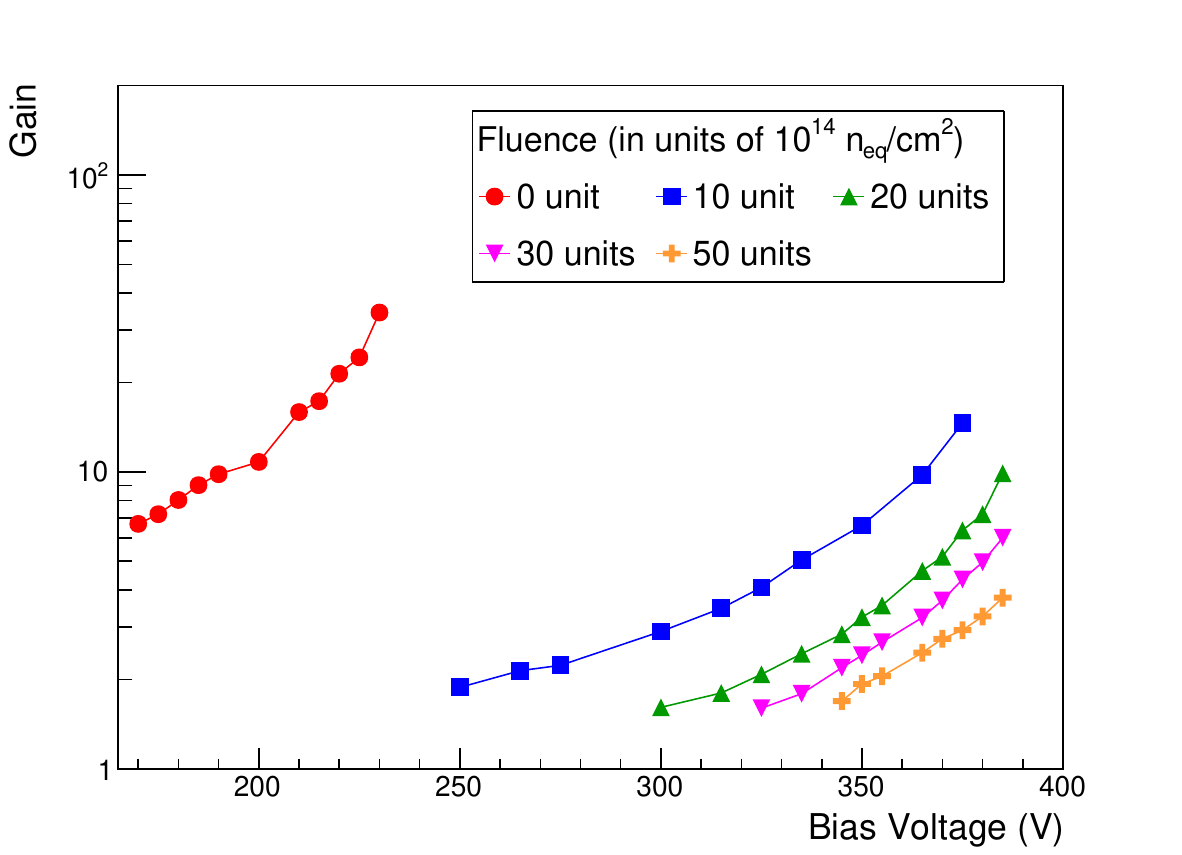}
    \caption{Simulated variation of gain of 20 $\upmu$m SiC bulk AC-LGAD over a range of operating bias voltage at different fluence levels.}
    \label{fig:9}
\end{figure}

Figure \ref{fig:9}  shows that, highly irradiated sensors have much lower gain values for the same bias voltage for a SiC bulk sensor. At a fluence of $10 \times 10^{14}\ \mathrm{n_{eq}~cm^{-2}}$, the sensor requires a bias voltage of 350 V to maintain a gain value same as that of a non-irradiated sensor operating at a bias voltage of 150 V. At higher fluence, a comparatively higher \(V_{bias}\) is required to compensate for the radiation-induced degradation and maintain the same gain.\par
It may also be observed that the change in voltage needed to maintain the same gain at high fluence is small in Fig~\ref{fig:9}. This behavior follows from the exponential acceptor-removal model, described in section~\ref{Simulation methodology}. At high fluence, further increases in fluence produce smaller changes in doping, so the bias needed to recover the same gain becomes smaller. This explains why the bias curves move closer together at high fluence in Fig~\ref{fig:9}.

In this study it was also noted that irradiation does not remove the gain layer abruptly. Instead, the peak doping decreases and the multiplication region becomes wider with fluence. For example, at 350\,V bias the peak electric field at zero fluence was found to be $\approx 452~\mathrm{kV/cm}$ with a peak acceptor concentration of $\approx 3.7\times10^{16}~\mathrm{cm^{-3}}$, while at a fluence of $5\times10^{15}~\mathrm{n_{eq}~cm^{-2}}$ the peak field decreases to $\approx 201~\mathrm{kV/cm}$ and the peak acceptor concentration to $\approx 2.4\times10^{15}~\mathrm{cm^{-3}}$. The field remains strong enough for multiplication at high fluence.

\subsubsection{Time resolution of irradiated sensors}

\begin{figure} 
    \centering
    \includegraphics[width=\linewidth]{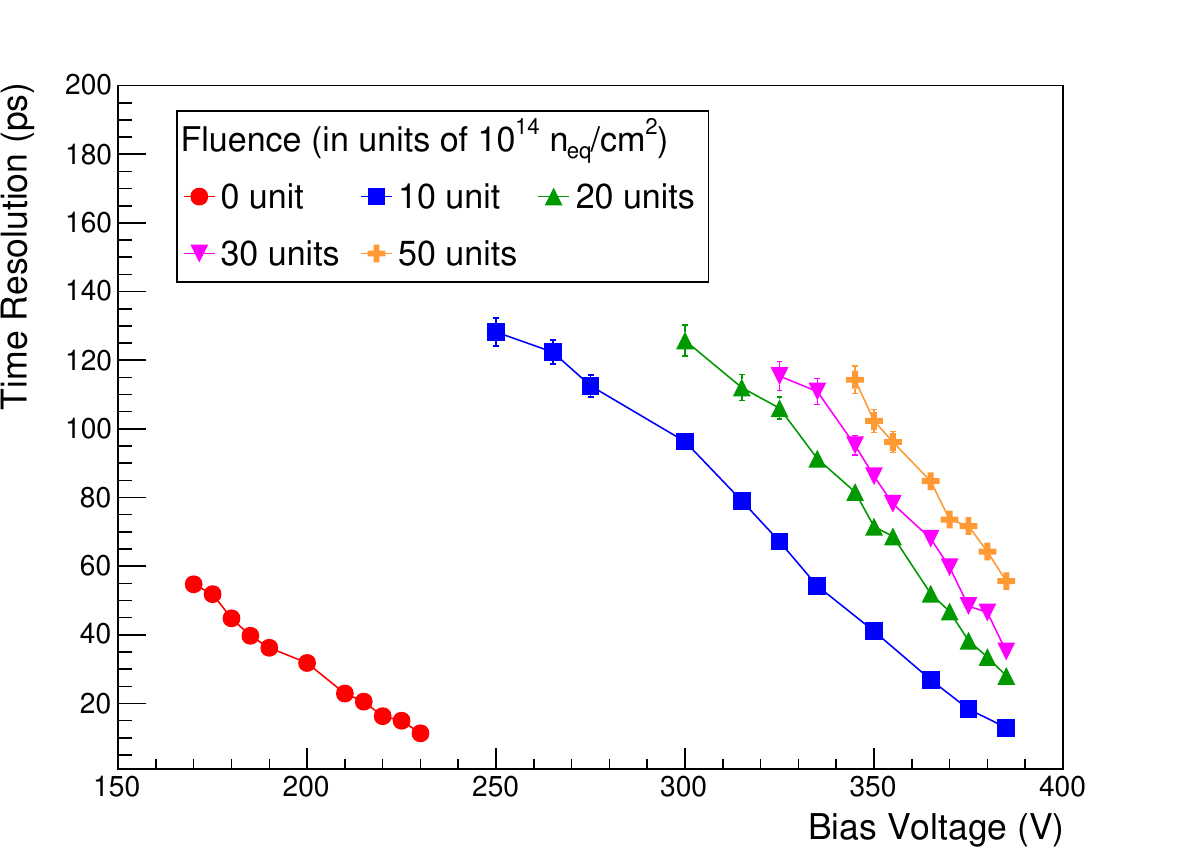}
    \caption{WF2 simulated time resolution of 20 $\upmu$m SiC bulk AC-LGAD with varying operational bias voltage at different irradiations.}
    \label{fig:10}
\end{figure}

The timing capability of an irradiated AC-LGAD sensor is studied and compared with that of a non-irradiated one. For this study, again an ultra-thin AC-LGAD with a 20 $\upmu$m SiC bulk, operated at 243 K is chosen. Figure \ref{fig:10} shows the time resolution  against bias voltage for different  fluences. The fluence range spans from $10 \times 10^{14}\ \mathrm{n_{eq}~cm^{-2}}$ to $50 \times 10^{14}\ \mathrm{n_{eq}~cm^{-2}}$, along with the non-irradiated case.

The red data points in Figure \ref{fig:10}, represents the non-irradiated sensor, which achieves a time resolution  below 10 ps. This represents an idealized case, assuming uniform electron-hole pair generation and excluding any random fluctuations in the simulation. The degradation in timing performance is evident when compared to the irradiated sample. Even at relatively low fluence, like $10 \times 10^{14}\ \mathrm{n_{eq}~cm^{-2}}$, there is a noticeable temporal shift, resulting in a clear distinction in timing performance. This degradation is primarily attributed to the acceptor removal mechanism in the gain layer, which reduces internal gain and lowers the signal-to-noise ratio.\cite{tcad}

For higher fluences, the curves shift towards higher bias voltages in order to achieve improved time resolution. Interestingly, the effect of irradiation appears less pronounced at higher fluences. This can be attributed to the initially higher concentration of boron dopants, which gradually become deactivated as radiation increases, ultimately leaving fewer active atoms susceptible to further effects.

A significant improvement in timing performance is observed as the bias increases from 300 V to 385 V. The most significant changes take place in this range, where the AC-LGAD's ability to resolve time is optimised. At the fluence \(20 \times 10^{14}~\mathrm{n_{eq}~cm^{-2}}\) and \(50 \times 10^{14}~\mathrm{n_{eq}~cm^{-2}}\), the time resolution improves by $\sim$77\% and $\sim$51\% within the voltage difference of $\Delta V = 85$ V and $\Delta V = 40$ V, respectively. The time resolution of less than 20 ps is achieved beyond 375 V for fluence \(50 \times 10^{14}~\mathrm{n_{eq}~cm^{-2}}\). This demonstrates how, and to what extent, the timing performance can be recovered in a moderately doped AC-LGAD after irradiation.

\begin{figure} 
    \centering
    \includegraphics[width=\linewidth]{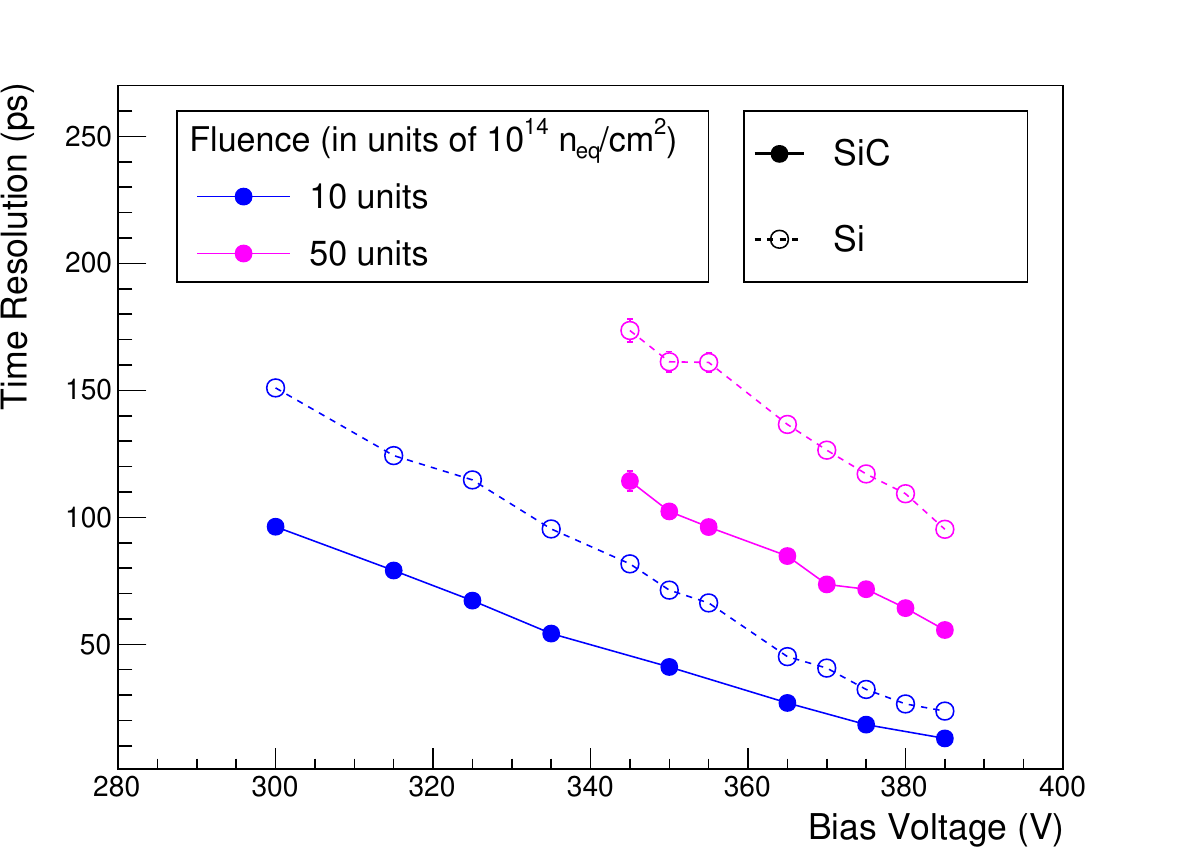}
    \caption{Comparison of WF2 simulated time resolution between a 20 $\upmu$m SiC and a 20 $\upmu$m Si bulk AC-LGAD with varying operational bias voltage.}
    \label{fig:11}
\end{figure}

In addition, the study of radiation tolerance is extended for a comparison with Si bulk AC-LGAD for $10 \times 10^{14}~\mathrm{n_{eq}~cm^{-2}}$ and $50 \times 10^{14}~\mathrm{n_{eq}~cm^{-2}}$. Figure \ref{fig:11} shows, the SiC based AC-LGAD sensors consistently outperform the Si one across the bias range. At a bias voltage of 365 V, for a fluence of $10 \times 10^{14}~\mathrm{n_{eq}~cm^{-2}}$, SiC bulk achieves a $\sim$ 40\% improvement in time resolution as compared to Si. At higher fluence of $50 \times 10^{14}~\mathrm{n_{eq}~cm^{-2}}$, improvement remains substantial with a $\sim$ 37\% reduction in time resolution value. This behaviour arises primarily from the higher carrier saturation velocity of silicon carbide compared to silicon, which directly enhances timing performance. In addition, the inherently higher atomic displacement energy of SiC, although not explicitly modelled in the present simulation, provides superior resistance to radiation-induced damage in real materials \cite{DeNapoli:2022}. The enhanced performance of the SiC bulk is therefore expected to be even more pronounced at high fluences.


\section{Conclusion}
This study presents a comprehensive investigation of ultra-thin LGAD sensors using the WeightField2 package. Data from FBK fabricated UFSD devices (silicon (DC) LGAD) are used to compare with WF2 simulation predictions for an identical device and found to be remarkably matching. This validates the effectiveness of WF2 in simulating LGAD devices. In the present study we assumed that the gain layer in 4H-SiC evolves under irradiation in the same way as in silicon. This assumption was necessary due to the lack of available data for irradiated 4H-SiC and in order to compare the two materials within the same simulation framework. We note that this has not yet been verified experimentally and should be considered an open subject for future studies.\par

It is observed that reducing the sensor thickness from 100~$\upmu$m to 20~$\upmu$m, at constant gain values, significantly improves the time resolution by $\sim$ 60\%. Furthermore, a simulated SiC sensor achieves around $\sim$ 15 ps better timing performance compared to a silicon device with same gain value (=10) and same (very low) thicknesses.

Temperature studies conducted on 20~$\upmu$m unirradiated SiC sensors show an increase in gain with decreasing temperature,  due to increasing ionisation probability at lower temperatures. This increase in gain is further boosted by higher bias voltages, where the time resolution also improves at lower temperatures and higher bias, reaching as low as $\sim$ 13~ps at 360~V. Furthermore, when compared to conventional silicon LGAD sensors, the SiC sensors shows a significant improvement in time resolution of $\sim$ 40\%, with decreasing temperature, emphasizing the superior timing response of SiC-based devices.

Radiation tolerance studies confirm significant differences in gain and time resolution between unirradiated sensors and irradiated sensors. The time resolution deteriorates by  $\sim$ 88\% for minimum to maximum fluence (irradiated case). The effect is reduced for higher bias voltages, and a time resolution of $\sim$  55 ps could be achieved for SiC bulk AC-LGAD at highest fluence by optimising operational parameters. Similar comparative study for the minimum and maximum fluence levels for Si bulk LGAD shows a deterioration of $\sim$ 72\% in time resolution in Si bulk at highest fluence when compared to SiC.\par

This paper has further shown the feasibility of significant enhancements in the gain and the time resolution at optimised sensor thickness and operating temperatures, with robust performance maintained even under a high radiation regime. Having already shown better performance of SiC LGAD device over a conventional Si LGAD device, we conclude that a fabricated ultra-thin SiC UFSD will demonstrate considerable superiority over Si sensors. This study aims to present a benchmark for comparison with future SiC AC-LGAD devices fabricated for 4D tracking applications under extreme conditions. 


\section*{Acknowledgments}
The authors gratefully acknowledge the Weightfield2 simulation package and sincerely thank its developers for making it freely available to the scientific community.
P. Palni and J. Kalani would also like to thank the SPS local cluster facility and the IIT Mandi SRIC seed grant support (Ref. No. IITM/SG/PP/128), they acknowledge the National Supercomputing Mission (NSM) for providing computing resources of ‘PARAM Himalaya’ at IIT Mandi, which is implemented by C-DAC and supported by the Ministry of Electronics and Information Technology (MeitY) and the Department of Science and Technology (DST), Government of India. S. Datta and G. J. Tambave acknowledge the infrastructure support of NISER, which is an OCI under Homi Bhabha National Institute; they acknowledge the funding of the Department of Atomic Energy, Government of India.

\onecolumn
\section*{Appendix}\label{appendix}
\begin{table}[htbp] 
\centering
\caption{Simulation parameters and their values for the FBK device experimental verification. (Fig 2)}
\renewcommand{\arraystretch}{1.2}
\begin{tabular}{|p{1.2cm}|p{6cm}|p{6cm}|}
\hline
\textbf{Sr. No.} & \textbf{Parameter} & \textbf{Value} \\ \hline
1 & Bulk Material & Si \\ \hline
2 & Structure & DC-LGAD \\ \hline
3 & Number of strips & 1  \\ \hline
4 & Thickness & 55~\textmu m \\ \hline
5 & Gain Layer Implant (GL) & 0.5-1~\textmu m \\ \hline
6 & Gain Dopant & B + C \\ \hline
7 & Gain Layer Doping & $4.71\times10^{16}$~cm$^{-3}$ \\ \hline
8 & Acceptor Removal (Doping Removal) & Kept ON (For irradiated samples) \\ \hline
9 & No. of e$^-$/h$^+$ Produced & 75~/~\textmu m (Uniform distribution) \\ \hline
10 & Temperature & 253~K (For irradiated samples) and 293~K (For unirradiated samples) \\ \hline
11 & Fluence & Varied as per reference study \\ \hline

\end{tabular}
\label{fbk_table}
\end{table}

\begin{table}[htbp]
\centering
\caption{Simulation parameters and their values used in the  study of choice of bulk material. (Fig 3)}
\renewcommand{\arraystretch}{1.2}
\begin{tabular}{|p{1.2cm}|p{6cm}|p{6cm}|}
\hline
\textbf{Sr. No.} & \textbf{Parameter} & \textbf{Value} \\ \hline
1 & Detector Type & AC-LGAD \\ \hline
2 & Bulk Material & SiC / Si / Diamond \\ \hline
3 & Number of strips & 3 \\ \hline
4 & Detector Thickness & 20 $\upmu$m \\ \hline
5 & Gain Layer Implant (GL) & 0.5-1~\textmu m \\ \hline
6 & Gain Doping & 4.3 (for SiC), 4.422 (for Si), 4.43 (for Diamond) (in units of $10^{16}$ n$_{eq}$\ cm$^{-3}$ \\ \hline
7 & Bias Voltage & 150 V \\ \hline
8 & No. of e$^-$/h$^+$ Produced & 57~/~\textmu m (SiC), 75~/~\textmu m (Si), 40~/~\textmu m (Diamond) (Uniform ionization) \\ \hline
9 & Temperature & 243~K\\ \hline
10 & Trapping Coefficients & $\beta_e = 4.9\times10^{-16}$~cm$^2$/ns, $\beta_h = 6.2\times10^{-16}$~cm$^2$/ns \\ \hline

\end{tabular}
\label{tab:thickness_study_parameters}
\end{table}

\begin{table}[htbp]
\centering
\caption{Simulation parameters and their values used in the thickness study for SiC AC-LGADs. (Fig 4)}
\renewcommand{\arraystretch}{1.2}
\begin{tabular}{|p{1.2cm}|p{6cm}|p{6cm}|}
\hline
\textbf{Sr. No.} & \textbf{Parameter} & \textbf{Value} \\ \hline
1 & Detector Type & AC-LGAD \\ \hline
2 & Bulk Material & SiC \\ \hline
3 & Number of strips & 5 \\ \hline
4 & Gain Layer Implant (GL) & 0.5-1~\textmu m \\ \hline
5 & Gain Doping & Varied concentrations of Boron \\ \hline
6 & Bias Voltage & 300~V \\ \hline
7 & No. of e$^-$/h$^+$ Produced & 57~/~\textmu m (SiC, Uniform ionization) \\ \hline
8 & Temperature & 243~K (Common simulation temperature) \\ \hline

\end{tabular}
\label{tab:thickness_study_parameters}
\end{table}

\begin{table}[htbp]
\centering
\caption{Simulation parameters and their settings used in the thickness study for Si and SiC AC-LGADs, gain comparison. (Fig 5,6)}
\renewcommand{\arraystretch}{1.2}
\begin{tabular}{|p{1.2cm}|p{6cm}|p{6cm}|}
\hline
\textbf{Sr. No.} & \textbf{Parameter} & \textbf{Value} \\ \hline
1 & Detector Type & AC-LGAD \\ \hline
2 & Bulk Material & Si / SiC \\ \hline
3 & Number of strips & 5 \\ \hline
4 & Thickness & 20-100~\textmu m (Different active layer thicknesses simulated) \\ \hline
5 & Gain Layer Implant (GL) & 0.5-1~\textmu m \\ \hline
6 & Gain Doping & $\sim(2.8$-$5.0)\times10^{16}~\mathrm{cm^{-3}}$ Boron atoms \\ \hline
7 & Bias Voltage & 300~V \\ \hline
8 & No. of e$^-$/h$^+$ Produced & 57~/~\textmu m (SiC), 75~/~\textmu m (Si) (Uniform ionization) \\ \hline
9 & Temperature & 243~K (Common simulation temperature) \\ \hline

\end{tabular}
\label{tab:thickness_study_parameters}
\end{table}

\begin{table}[htbp]
\centering
\caption{Simulation parameters and their values used in the temperature study for SiC AC-LGAD gain. (Fig 7)}
\renewcommand{\arraystretch}{1.2}
\begin{tabular}{|p{1.2cm}|p{6cm}|p{6cm}|}
\hline
\textbf{Sr. No.} & \textbf{Parameter} & \textbf{Value} \\ \hline
1 & Detector Type & AC-LGAD \\ \hline
2 & Bulk Material & SiC \\ \hline
3 & Number of strips & 5 \\ \hline
4 & Thickness & 20~\textmu m (ultra-thin) \\ \hline
5 & Gain Layer Implant (GL) & 0.5-1~\textmu m \\ \hline
6 & Gain Doping & $2.301\times10^{16}$ cm$^{-3}$ (Boron atoms) \\ \hline
7 & No. of e$^-$/h$^+$ Produced & 57~/~\textmu m (SiC, Uniform ionization) \\ \hline

\end{tabular}
\label{tab:thickness_study_parameters}
\end{table}

\begin{table}[htbp]
\centering
\caption{Simulation parameters and their values used in the temperature study for Si and SiC AC-LGADs, time resolution comparison. (Fig 8)}
\renewcommand{\arraystretch}{1.2}
\begin{tabular}{|p{1.2cm}|p{6cm}|p{6cm}|}
\hline
\textbf{Sr. No.} & \textbf{Parameter} & \textbf{Value} \\ \hline
1 & Detector Type & AC-LGAD \\ \hline
2 & Bulk Material & SiC / Si \\ \hline
3 & Number of strips & 5 \\ \hline
4 & Thickness & 20~\textmu m (ultra-thin) \\ \hline
5 & Gain Layer Implant (GL) & 0.5-1~\textmu m \\ \hline
6 & Gain layer doping & $2.301\times10^{16}$ cm$^{-3}$ (Boron atoms) \\ \hline
7 & No. of e$^-$/h$^+$ Produced & 57~/~\textmu m (SiC), 75~/~\textmu m (Si) (Uniform ionization) \\ \hline

\end{tabular}
\label{tab:thickness_study_parameters}
\end{table}

\begin{table}[htbp]
\centering
\caption{Simulation parameters and their values used in the thickness-fluence study of SiC AC-LGADs. (Fig 9)}
\renewcommand{\arraystretch}{1.2}
\begin{tabular}{|p{1.2cm}|p{6cm}|p{6cm}|}
\hline
\textbf{Sr. No.} & \textbf{Parameter} & \textbf{Value} \\ \hline
1 & Detector Type & AC-LGAD \\ \hline
2 & Bulk Material & SiC \\ \hline
3 & Number of strips & 1 \\ \hline
4 & Thickness & 20-120~\textmu m \\ \hline
5 & Gain Layer Implant (GL) & 0.5-1~\textmu m \\ \hline
6 & Gain layer doping & $4.40-4.94\times10^{16}$ cm$^{-3}$ (Boron atoms) \\ \hline
7 & No. of e$^-$/h$^+$ Produced & 57~/~\textmu m uniform ionisation for SiC \\ \hline
8 & Temperature & 253 K \\ \hline
9 & Bias Voltage & 115-200 V \\ \hline

\end{tabular}
\label{tab:thickness_study_parameters}
\end{table}

\begin{table}[htbp]
\centering
\caption{Simulation parameters and their values used in the radiation hardness study for SiC AC-LGADs and comparison with Si (Fig.10, 11, 12).}
\renewcommand{\arraystretch}{1.2}
\begin{tabular}{|p{1.2cm}|p{6cm}|p{6cm}|}
\hline
\textbf{Sr. No.} & \textbf{Parameter} & \textbf{Value} \\ \hline
1 & Detector Type & AC-LGAD \\ \hline
2 & Bulk Material & SiC / Si \\ \hline
3 & Number of Strips & 5 \\ \hline
4 & Thickness & 20~\textmu m (Ultra-thin) \\ \hline
5 & Gain Layer Implant (GL) & 0.5-1~\textmu m \\ \hline
6 & Gain Doping & $3.701\times10^{16}$~cm$^{-3}$ (Boron doped) \\ \hline
7 & Acceptor Removal & Kept ON \\ \hline
8 & Trapping Coefficients & $\beta_e = 4.9\times10^{-16}$~cm$^2$/ns, $\beta_h = 6.2\times10^{-16}$~cm$^2$/ns \\ \hline
9 & Temperature & 243~K (Common) \\ \hline
10 & No. of e$^-$/h$^+$ Produced & 57~/~\textmu m (SiC), 75~/~\textmu m (Si) (Uniform ionization) \\ \hline

\end{tabular}
\label{tab:radiation_study_parameters}
\end{table}

\clearpage
\twocolumn

\end{document}